\documentclass[12pt,a4paper]{article}
\usepackage{amsmath,amssymb}
\usepackage[sort]{cite}
\newcommand{\ind}[2]{^{#1}_{\text{#2}}}
\def\Nc{N_{\text c}}
\def\nf{n_{f}}

\begin{document}

\begin{center}

{\Large\bf Recurrent form of the renormalization group relations
for the higher--order hadronic vacuum polarization function
perturbative expansion coefficients

}

\vskip10mm

{\large A.V.~Nesterenko}

\vskip7.5mm

{\small\it Bogoliubov Laboratory of Theoretical Physics,
Joint Institute for Nuclear Research,\\
Dubna, 141980, Russian Federation}

\end{center}

\vskip5mm

\noindent
\centerline{\bf Abstract}

\vskip2.5mm

\centerline{\parbox[t]{150mm}{%
The renormalization group relations for the higher--order hadronic vacuum
polarization function perturbative expansion coefficients are studied. The
folded recurrent and unfolded explicit forms of such relations are
obtained. The explicit expression for the coefficients, which incorporate
the contributions of the $\pi^{2}$--terms in the perturbative expansion of
the $R$--ratio of electron--positron annihilation into hadrons, is derived
at an arbitrary loop level. The obtained results can be employed as an
independent crosscheck of the higher--order perturbative calculations of
the hadronic vacuum polarization function and in the studies of the
renormalization scale setting in the relevant physical observables.
\\[2.5mm]
}}

\vskip12mm

\section{Introduction}
\label{Sect:Intro}

A variety of issues of elementary particle physics is inherently based on
the hadronic vacuum polarization function~$\Pi(q^2)$, the corresponding
Adler function~$D(Q^2)$, and the function~$R(s)$. In~particular, these
functions play a key role in the theoretical studies of numerous strong
interaction processes, for example, electron--positron annihilation into
hadrons, inclusive $\tau$~lepton hadronic decay, and hadronic
contributions to such precise electroweak observables as the muon
anomalous magnetic moment and the running of the electromagnetic coupling.
Basically, all these processes are of a straight relevance to the future
collider projects~\cite{FCC, CEPC, ILC, CLIC, E989, E34, MUonE},
contemporary experiments~\cite{ALEPH, OPAL, HFLAV, BES, KEDR}, and various
ongoing research programs, see, e.g.,~Refs.~\mbox{\cite{FCCtheory19,
MuonWP20}}. The~theoretical investigation of these issues provides a
decisive consistency test of the Standard Model and imposes firm
restrictions on a possible new physics beyond the latter. In~turn, an
anticipated increase of the accuracy of measurements at the future
experimental facilities calls for a further refinement of theoretical
approaches, which address both perturbative and intrinsically
nonperturbative aspects of hadron dynamics.

Factually, the perturbation theory and the renormalization group~(RG)
method remain to be a basic tool for the theoretical exploration of
Quantum Chromodynamics~(QCD) at high energies. Specifically, the
calculation of pertinent Feynman diagrams constitutes a reliable way to
obtain the perturbative expression for the hadronic vacuum polarization
function of the form of Eq.~(\ref{PpertMu}) below. At~the same time, the
RG~method also enables one to tie together the higher--order
coefficients~$\Pi_{j,k}$ entering Eq.~(\ref{PpertMu}). At~the few lowest
orders of perturbation theory these RG~relations assume a rather simple
form, whereas at the higher orders such relations become quite cumbersome
and their derivation requires considerable efforts.

The~primary objective of the paper is to obtain the RG~relations for the
higher--order hadronic vacuum polarization function perturbative expansion
coefficients~$\Pi_{j,k}$ in a folded recurrent and unfolded explicit
forms. It~is also of an apparent interest to derive, at~an arbitrary loop
level, the explicit expression for the coefficients embodying the
\mbox{$\pi^{2}$--terms} in the perturbative expansion of the
\mbox{$R$--ratio} of electron--positron annihilation into hadrons, which
play a substantial role for the subject on~hand.

The layout of the paper is as follows. Section~\ref{Sect:Methods}
delineates the hadronic vacuum polarization function~$\Pi(q^2)$, the Adler
function~$D(Q^2)$, and the function~$R(s)$, recaps their perturbative
expressions, and expounds the corresponding RG~equations.
In~Sect.~\ref{Sect:Results} the RG~relations for the higher--order
hadronic vacuum polarization function perturbative expansion
coefficients~$\Pi_{j,k}$ are obtained in a folded recurrent and unfolded
explicit forms, the explicit expression for the coefficients, which
incorporate the \mbox{$\pi^{2}$--terms} in the perturbative expansion of
the \mbox{$R$--ratio} of electron--positron annihilation into hadrons, is
derived at~an arbitrary loop level, and the discussion of practical
applications of the obtained results is presented.
Section~\ref{Sect:Concl} summarizes the basic results.
Appendix~\ref{Sect:CoeffsRG} contains a~supplementary material.

\section{Methods}
\label{Sect:Methods}

\subsection{Functions~$\Pi(q^2)$, $D(Q^2)$, and~$R(s)$}
\label{Sect:PDRgen}

As~discussed earlier, the theoretical study of a broad pattern of the
strong interaction processes is inherently based on the hadronic vacuum
polarization function~$\Pi(q^2)$, which is defined as the scalar part of
the hadronic vacuum polarization tensor
\begin{equation}
\label{P_Def}
\Pi_{\mu\nu}(q^2) = i\!\int\!d^4x\,e^{i q x} \bigl\langle 0 \bigl|\,
T\!\left\{J_{\mu}(x)\, J_{\nu}(0)\right\} \bigr| 0 \bigr\rangle =
\frac{i}{12\pi^2} (q_{\mu}q_{\nu} - g_{\mu\nu}q^2) \Pi(q^2),
\end{equation}
with $q^2<0$~being the spacelike kinematic variable. The
function~$\Pi(q^2)$~(\ref{P_Def}) satisfies the inhomogeneous RG~equation
\begin{equation}
\label{RGeqnPgen}
\left[\frac{\partial}{\partial\ln\mu^2} +
\frac{\partial a\ind{}{s}(\mu^2)}{\partial\ln\mu^2}
\frac{\partial}{\partial a\ind{}{s}}\right]\!
\Pi\bigl(q^2,\mu^2,a\ind{}{s}\bigr) = \gamma(a\ind{}{s}),
\end{equation}
where $\mu^2>0$~denotes the renormalization scale and~$\gamma(a\ind{}{s})$
stands for the anomalous dimension. In~Eq.~(\ref{RGeqnPgen})
$a\ind{}{s}(\mu^2) = \alpha\ind{}{s}(\mu^2)\beta_{0}/(4\pi)$ is the
\mbox{so--called} QCD~couplant, which satisfies the RG~equation
\begin{equation}
\label{RGeqnAgen}
\frac{\partial a\ind{}{s}(\mu^2)}{\partial\ln\mu^2} =
\beta(a\ind{}{s}),
\end{equation}
where~$\beta(a\ind{}{s})$ denotes the $\beta$~function, $\beta_{0} = 11 -
2\nf/3$ is the one--loop $\beta$~function perturbative expansion
coefficient, and $\nf$~stands for the number of active flavors. For
practical applications it is particularly convenient to deal with the
Adler function, which is defined as the logarithmic derivative of the
hadronic vacuum polarization function~\cite{Adler}
\begin{equation}
\label{Adler_Def}
D(Q^2) = - \frac{d\, \Pi(-Q^2)}{d \ln Q^2},
\end{equation}
with~$Q^2=-q^2>0$ being the spacelike kinematic variable. The Adler
function satisfies the homogeneous RG~equation
\begin{equation}
\label{RGeqnDgen}
\left[\frac{\partial}{\partial\ln\mu^2} +
\frac{\partial a\ind{}{s}(\mu^2)}{\partial\ln\mu^2}
\frac{\partial}{\partial a\ind{}{s}}\right]\!
D\bigl(Q^2,\mu^2,a\ind{}{s}\bigr) = 0.
\end{equation}

In~turn, the function~$R(s)$, which is commonly identified with the
$R$--ratio of electron--positron annihilation into hadrons $R(s) =
\sigma(e^{+}e^{-} \!\to \text{hadrons}; s)/\sigma(e^{+}e^{-} \!\to
\mu^{+}\mu^{-}; s)$, can be calculated as the discontinuity of the
hadronic vacuum polarization function across the physical~cut
\begin{equation}
\label{R_Def}
R(s) = \frac{1}{2 \pi i} \lim_{\varepsilon \to 0_{+}}
\Bigl[\Pi(s+i\varepsilon) - \Pi(s-i\varepsilon)\Bigr]\! =
\frac{1}{\pi}\,{\rm Im} \!\lim_{\varepsilon \to 0_{+}}\!
\Pi(s+i\varepsilon)
\end{equation}
or, equivalently, by integrating Eq.~(\ref{Adler_Def}) in finite limits,
specifically~\cite{Rad82, KP82}
\begin{equation}
\label{R_Disp2}
R(s) =  \frac{1}{2 \pi i} \lim_{\varepsilon \to 0_{+}}
\int\limits_{s + i \varepsilon}^{s - i \varepsilon}
D(-\zeta)\,\frac{d \zeta}{\zeta},
\end{equation}
where \mbox{$s=q^2>0$} denotes the timelike kinematic variable, namely,
the center--of--mass energy squared. In~Eq.~(\ref{R_Disp2}) the
integration contour in the complex $\zeta$--plane lies in the region of
analyticity of the integrand. The~complete set of relations, which express
the functions on hand in terms of each other, as well as a detailed
discussion of their physical implications can be found in, e.g.,
Ref.~\cite{Book} and references therein.

\subsection{Functions~$\Pi(q^2)$, $D(Q^2)$, and~$R(s)$ within perturbation theory}
\label{Sect:PDRpert}

In the framework of perturbation theory the $\ell$--loop expression for
the hadronic vacuum polarization function~(\ref{P_Def}) can be
represented~as
\begin{equation}
\label{PpertMu}
\Pi^{(\ell)}\bigl(q^2,\mu^2,a\ind{}{s}\bigr) =
\sum_{j=0}^{\ell}\Bigl[a\ind{(\ell)}{s}(\mu^2)\Bigr]^{j}
\sum_{k=0}^{j+1}\Pi_{j,k}\ln^{k}\!\!\left(\frac{\mu^2}{-q^2}\right)\!,
\qquad
q^2 \to -\infty
\end{equation}
and, as~immediatedly follows from Eq.~(\ref{Adler_Def}), the corresponding
expression for the Adler function~reads
\begin{equation}
\label{DpertMu}
D^{(\ell)}\bigl(Q^2,\mu^2,a\ind{}{s}\bigr) =
\sum_{j=0}^{\ell}\Bigl[a\ind{(\ell)}{s}(\mu^2)\Bigr]^{j}
\sum_{k=0}^{j+1}k\Pi_{j,k}\ln^{k-1}\!\!\left(\frac{\mu^2}{Q^2}\right)\!,
\qquad
Q^2 \to \infty.
\end{equation}
Note that the~common prefactor $\Nc\sum_{f=1}^{\nf} Q_{f}^{2}$ is omitted
throughout, where \mbox{$\Nc=3$}~denotes the number of colors and
$Q_{f}$~stands for the electric charge of $f$--th quark. The native choice
of the renormalization scale~$\mu^2=Q^2$ (that amounts to the RG~summation
in the spacelike domain) casts Eq.~(\ref{DpertMu}) to a well--known
form~($\Pi_{0,1}=1$)
\begin{equation}
\label{DpertDef}
D^{(\ell)}(Q^2) =
\sum_{j=0}^{\ell}\Pi_{j,1}\Bigl[a\ind{(\ell)}{s}(Q^2)\Bigr]^{j} =
1 + \sum_{j=1}^{\ell} d_{j} \Bigl[a\ind{(\ell)}{s}(Q^2)\Bigr]^{j}\!,
\qquad
d_{j} =\Pi_{j,1},
\end{equation}
whereas the elimination of dependence of Eq.~(\ref{PpertMu}) on the
renormalization scale was discussed in, e.g., Refs.~\cite{RPert1L, Penn,
Pivovarov91, JPG42, JPG46, Book}. It~is~worthwhile to mention here that a
general choice of the renormalization scale~$\mu^2=c\,Q^2$ (with~$c \neq
1$ being a positive constant) keeps in the resulting expression for the
Adler function~$D^{(\ell)}(Q^2)$ all the terms proportional to the
higher--order coefficients~$\Pi_{j,k}$ ($0 \le j \le \ell$, $0 \le k \le
j+1$) appearing on the right--hand side of~Eq.~(\ref{DpertMu}). In~turn,
at~the $\ell$--loop level the anomalous dimension and $\beta$~function
entering Eqs.~(\ref{RGeqnPgen}) and~(\ref{RGeqnDgen}) take the
following~form
\begin{equation}
\label{GpertDef}
\gamma^{(\ell)}\bigl(a\ind{}{s}\bigr) =
\sum_{j=0}^{\ell}\gamma_{j}\Bigl[a\ind{(\ell)}{s}(\mu^2)\Bigr]^{j},
\end{equation}
\begin{equation}
\label{BpertDef}
\beta^{(\ell)}\bigl(a\ind{}{s}\bigr) =
-\sum_{i=0}^{\ell-1}B_{i}\Bigl[a\ind{(\ell)}{s}(\mu^2)\Bigr]^{i+2},
\qquad
B_{i} = \frac{\beta_{i}}{\beta_{0}^{i+1}}.
\end{equation}
The explicit expressions for the perturbative expansion
coefficients~$\Pi_{j,k}$~(\ref{PpertMu}) and~$\gamma_{j}$~(\ref{GpertDef})
are currently available up to the fourth order in the QCD
couplant~$a\ind{}{s}$, whereas the
coefficients~$\beta_{i}$~(\ref{BpertDef}) are known up to the fifth order
in~$a\ind{}{s}$, see Refs.~\cite{RPert4L, BCK0912} and~\cite{Beta5L},
respectively.

As~for the function~$R(s)$, it can be calculated by making use of either
of Eqs.~(\ref{R_Def}) and~(\ref{R_Disp2}). At~the same time, it is
necessary to emphasize here that the RG~summation must be performed in
Eqs.~(\ref{PpertMu}) and~(\ref{DpertMu}) in the spacelike domain prior to
the application of Eqs.~(\ref{R_Def}) and~(\ref{R_Disp2}), respectively.
Otherwise the effects due to continuation of theoretical results from
spacelike into timelike domain may not be properly accounted~for.
In~particular, the direct calculation of discontinuity~(\ref{R_Def}) of
expression~(\ref{PpertMu}) with subsequent assignment of the
renormalization scale~\mbox{$\mu^2=|s|$} (that factually amounts to an
incomplete RG~summation in the timelike domain, see, e.g.,
Refs.~\cite{Penn, Pivovarov91, EPJC77, JPG46}) yields
\begin{equation}
\label{Rappr}
R^{(\ell)}(s) =
1 + \sum_{j=1}^{\ell} r_{j} \Bigl[a\ind{(\ell)}{s}(|s|)\Bigr]^{j},
\qquad
r_{j} = d_{j} - \delta_{j},
\end{equation}
where $d_{j} =\Pi_{j,1}$ stand for the Adler function perturbative
expansion coefficients~(\ref{DpertDef}) and $\delta_{j}$~embody the
contributions of the so--called~$\pi^2$--terms (which play a significant
role in the studies of the process on~hand), namely
\begin{equation}
\label{DeltaGen}
\delta_{1} = 0,
\qquad
\delta_{2} = 0,
\qquad
\delta_{j} = \sum_{k=1}^{K(j)} (-1)^{k+1} \,\pi^{2k}\, \Pi_{j,2k+1},
\qquad
j \ge 3.
\end{equation}
In this equation
\begin{equation}
\label{KImDef}
K(n) = \frac{n-2}{2} + \frac{n \;\mbox{mod}\; 2}{2}
\end{equation}
and $(n \;\mbox{mod}\; m)$ is the remainder on division of~$n$ by~$m$.
It~is worthwhile to note also that the
function~$R^{(\ell)}(s)$~(\ref{Rappr}) only partially retains the effects
due to continuation of theoretical results from spacelike into timelike
domain and has certain shortcomings, see, e.g., Refs.~\cite{Penn, Rad82,
KP82, Bj89, Pivovarov91, ProsperiAlpha, EPJC77, JPG46, Book} and
references therein for a detailed discussion of these issues.

\section{Results and discussion}
\label{Sect:Results}

\subsection{Recurrence relations for the coefficients~$\Pi_{j,k}$}
\label{Sect:RR}

As~mentioned earlier, the RG equations for the functions on hand bind
together the higher--order perturbative coefficients~$\Pi_{j,k}$ appearing
in Eqs.~(\ref{PpertMu}) and~(\ref{DpertMu}). In~particular, at an
arbitrary loop level the coefficients~$\Pi_{j,k}$ $(j \ge 1; \,
k=1,\ldots,j+1)$ can be expressed in terms of the
coefficients~$\gamma_{i}$ $(i=1,\ldots,j)$ entering Eq.~(\ref{GpertDef})
and (if~$j \ge 2$) $\Pi_{i,0}$~$(i=1,\ldots,j-1)$. Specifically,
Eqs.~(\ref{RGeqnPgen}), (\ref{RGeqnAgen}), (\ref{PpertMu}),
(\ref{GpertDef}), and~(\ref{BpertDef}) imply that
\begin{align}
\label{RGeqnP}
& \sum_{j=0}^{\ell}\Bigl[a\ind{(\ell)}{s}(\mu^2)\Bigr]^{j}\!
\left\{\!\left[
\sum_{k=0}^{j+1}k\Pi_{j,k}\ln^{k-1}\!\!\left(\frac{\mu^2}{-q^2}\right)\!
\right]\! - \gamma_{j} \right\}
= \nonumber \\[2mm]
& = \left\{
\sum_{j=0}^{\ell}j\Bigl[a\ind{(\ell)}{s}(\mu^2)\Bigr]^{j-1}
\sum_{k=0}^{j+1}\Pi_{j,k}\ln^{k}\!\!\left(\frac{\mu^2}{-q^2}\right)\!\!
\right\}
\left\{
\sum_{i=0}^{\ell-1}B_{i}\Bigl[a\ind{(\ell)}{s}(\mu^2)\Bigr]^{i+2}
\right\}\!.
\end{align}
For~example, at the first few orders of perturbation theory this equation
yields $\Pi_{0,1} = \gamma_{0}$, $\Pi_{1,1} = \gamma_{1}$, $\Pi_{2,1} =
\Pi_{1,0} + \gamma_{2}$, \mbox{$\Pi_{2,2} = \gamma_{1}/2$}
($\Pi_{j,j+1}=0$ for~$j \ge 1$), see, e.g., Refs.~\cite{BCK0912, JPG46},
whereas at the higher--loop levels the corresponding RG~relations for the
coefficients~$\Pi_{j,k}$ become quite tangled and can be found in
Appendix~A of Ref.~\cite{JPG46}.

At~the same time, the RG~equation for the Adler function~(\ref{RGeqnDgen})
provides the relations for the higher--order perturbative expansion
coefficients~$\Pi_{j,k}$, which, being equivalent to the aforementioned
ones, have a somewhat different form. Namely, at any given loop level
Eqs.~(\ref{RGeqnDgen}), (\ref{RGeqnAgen}), (\ref{DpertMu}),
and~(\ref{BpertDef}) enable one to express the coefficients~$\Pi_{j,k}$
$(j \ge 2; \, k=2,\ldots,j+1)$ in terms of the
coefficients~$\Pi_{i,1}$~$(i=1,\ldots,j-1)$, specifically
\begin{align}
\label{RGeqnD}
& \sum_{j=0}^{\ell}\Bigl[a\ind{(\ell)}{s}(\mu^2)\Bigr]^{j}\!
\left[
\sum_{k=0}^{j+1}k(k-1)\Pi_{j,k}\ln^{k-2}\!\!\left(\frac{\mu^2}{Q^2}\right)\!
\right] \!
= \nonumber \\[2mm]
& = \left\{
\sum_{j=0}^{\ell}j\Bigl[a\ind{(\ell)}{s}(\mu^2)\Bigr]^{j-1}
\sum_{k=0}^{j+1}k\Pi_{j,k}\ln^{k-1}\!\!\left(\frac{\mu^2}{Q^2}\right)\!\!
\right\}
\left\{
\sum_{i=0}^{\ell-1}B_{i}\Bigl[a\ind{(\ell)}{s}(\mu^2)\Bigr]^{i+2}
\right\}\!.
\end{align}
In~particular, at the first few orders of perturbation theory
Eq.~(\ref{RGeqnD}) yields $\Pi_{2,2} = \Pi_{1,1}/2$, $\Pi_{3,2} =
\Pi_{1,1}(B_{1}/2) + \Pi_{2,1}$, $\Pi_{3,3} = \Pi_{1,1}/3$
($\Pi_{j,j+1}=0$ for~$j \ge 1$), whereas the RG relations for the
coefficients~$\Pi_{j,k}$ at the higher orders become rather cumbrous and
are given in Appendix~\ref{Sect:CoeffsRG}.

At~an arbitrary loop level the RG~relations for the
coefficients~$\Pi_{j,k}$, which follow from Eq.~(\ref{RGeqnD}), can be
represented in a compact recurrent form, namely
\begin{equation}
\label{PIj2Rec}
\Pi_{j,2} = \frac{1}{2} \sum_{i=1}^{j-1} i\, B_{j-i-1} \Pi_{i,1},
\qquad
j \ge 2,
\end{equation}
\begin{equation}
\label{PIjkRec}
\Pi_{j,k} = \frac{1}{T_{k-1}} \sum_{i=k-2}^{j-2}
i\, (i+j) \mathfrak{B}_{j-i-2} \Pi_{i,k-2},
\qquad
j \ge k,
\qquad
k \ge 3,
\end{equation}
where
\begin{equation}
\label{CBdef}
\mathfrak{B}_{n} = \frac{1}{4} \sum_{i=0}^{n} B_{i}\,B_{n-i},
\qquad
B_{i} = \frac{\beta_{i}}{\beta_{0}^{i+1}}
\end{equation}
and
\begin{equation}
T_{n} = \frac{1}{2}\, n (n+1)
\end{equation}
denotes the so--called triangular number.

The~obtained recurrence relations~(\ref{PIj2Rec}) and~(\ref{PIjkRec}) can
also be unfolded, that eventually results in the following explicit
expressions for the higher--order coefficients~$\Pi_{j,k}$ for,
respectively, odd and even values of the second index:
\begin{align}
\label{PIjkOdd}
\Pi_{j,2k+1} = & \,\frac{2^{k}}{(2k+1)!}
\sum_{i_{1}=2(k-1)+1}^{j-2}
\underbrace{%
\sum_{i_{2}=2(k-2)+1}^{i_{1}-2}   \ldots
\sum_{i_{n}=2(k-n)+1}^{i_{n-1}-2} \ldots
\sum_{i_{k}=1}^{i_{k-1}-2}%
}_{\mbox{{\small$(k-1)$~sums}}}
(j+i_{1})i_{1} \times
\nonumber \\[1.5mm] & \times
\underbrace{%
(i_{1}+i_{2})i_{2} \times \ldots \times
(i_{n-1}+i_{n})i_{n} \times \ldots \times
(i_{k-1}+i_{k})i_{k}%
}_{\mbox{{\small$(k-1)$~products}}}
\times
\nonumber \\[2mm] & \times
\mathfrak{B}_{j-i_{1}-2}
\underbrace{%
\mathfrak{B}_{i_{1}-i_{2}-2} \ldots
\mathfrak{B}_{i_{n-1}-i_{n}-2} \ldots
\mathfrak{B}_{i_{k-1}-i_{k}-2}%
}_{\mbox{{\small$(k-1)$~terms}}}
\Pi_{i_{k},1},
\qquad
j \ge (2k+1),
\quad
k \ge 1,
\end{align}
\begin{align}
\label{PIjkEven}
\Pi_{j,2k} = & \,\frac{2^{k-1}}{(2k)!}
\sum_{i_{1}=2(k-1)}^{j-2}
\underbrace{%
\sum_{i_{2}=2(k-2)}^{i_{1}-2}   \ldots
\sum_{i_{n}=2(k-n)}^{i_{n-1}-2} \ldots
\sum_{i_{k-1}=2}^{i_{k-2}-2}%
}_{\mbox{{\small$(k-2)$~sums}}}
\sum_{i_{k}=1}^{i_{k-1}-1}
(j+i_{1})i_{1} \times
\nonumber \\[1.5mm] & \times
\underbrace{%
(i_{1}+i_{2})i_{2} \times \ldots \times
(i_{n-1}+i_{n})i_{n} \times \ldots \times
(i_{k-2}+i_{k-1})i_{k-1}%
}_{\mbox{{\small$(k-2)$~products}}}
\times i_{k} \times
\nonumber \\[2mm] & \times
\mathfrak{B}_{j-i_{1}-2}
\underbrace{%
\mathfrak{B}_{i_{1}-i_{2}-2} \ldots
\mathfrak{B}_{i_{n-1}-i_{n}-2} \ldots
\mathfrak{B}_{i_{k-2}-i_{k-1}-2}%
}_{\mbox{{\small$(k-2)$~terms}}}
B_{i_{k-1}-i_{k}-1} \Pi_{i_{k},1},
\quad
j \ge 2k,
\quad
k \ge 2,
\end{align}
where the coefficients~$\mathfrak{B}_{n}$ and~$B_{i}$ are defined in
Eq.~(\ref{CBdef}). It~is straightforward to verify that
Eqs.~(\ref{PIjkOdd}) and~(\ref{PIjkEven}) reproduce the RG~relations for
the coefficients~$\Pi_{j,k}$, which follow from Eq.~(\ref{RGeqnD}), see
Appendix~\ref{Sect:CoeffsRG}. It~is worthwhile to mention that in certain
cases Eqs.~(\ref{PIj2Rec}),~(\ref{PIjkOdd}), and~(\ref{PIjkEven}) can be
represented in a somewhat simpler form, in~particular,
\begin{equation}
\Pi_{j,j-1} = (H_{j-1}-1)B_{1}\Pi_{1,1} + \Pi_{2,1},
\quad
j \ge 3;
\qquad
\Pi_{j,j} = \frac{1}{j}\,\Pi_{1,1},
\quad
j \ge 2,
\end{equation}
where
\begin{equation}
H_{n} = \sum_{k=1}^{n}\frac{1}{k}
\end{equation}
stands for the so--called harmonic number.

At~the same time, the higher--order coefficients~$\Pi_{j,k}$ can also be
expressed in terms of the coefficients~$\Pi_{i,0}$
and~$\gamma_{i}$~(\ref{GpertDef}). Specifically, for this purpose the
obtained results should be supplemented by the relations~\cite{JPG46}
\begin{equation}
\label{PIj1Rels}
\Pi_{0,1} = \gamma_{0},
\quad\;\;
\Pi_{1,1} = \gamma_{1},
\quad\;\;
\Pi_{j,1} = \gamma_{j} + \sum_{k=1}^{j-1}k\,\Pi_{k,0}\,B_{j-k-1},
\quad\;\;
B_{j} = \frac{\beta_{j}}{\beta_{0}^{j+1}},
\quad\;\;
j \ge 2.
\end{equation}
In~particular, Eqs.~(\ref{PIj2Rec}), (\ref{PIjkOdd}), (\ref{PIjkEven}),
and~(\ref{PIj1Rels}), being applied jointly, reproduce the RG~relations
for the coefficients~$\Pi_{j,k}$, which follow from Eq.~(\ref{RGeqnP}),
see Appendix~A of Ref.~\cite{JPG46}.

\subsection{Discussion}
\label{Sect:Disc}

First of all, it is necessary to outline that the results obtained in
Sect.~\ref{Sect:RR} can be employed as an independent crosscheck for the
perturbative calculations at the higher loop levels. In~particular,
Eqs.~(\ref{PIj2Rec}), (\ref{PIjkOdd}),~(\ref{PIjkEven}) enable one to find
explicit expressions for all the coefficients~$\Pi_{j,k}$, which involve
the hitherto known perturbative coefficients~$\Pi_{i,1}$. As~mentioned
earlier, the latter are currently available up to the fourth order in the
strong running coupling~$(i=0,\ldots,4)$, therefore
Eqs.~(\ref{PIj2Rec}),~(\ref{PIjkOdd}),~(\ref{PIjkEven}) provide, at an
arbitrary loop level~(i.e.,~for any given~$j \ge 2$), the explicit
expressions for the higher--order
coefficients~$\Pi_{j,j-n}$~(\mbox{$n=0,\ldots,\rm{min}\{j-2,3\}$}).
For~example, at the~10th and~25th orders of perturbation~theory
\begin{equation}
\Pi_{10,9} =
\frac{4609}{2520}\, B_{1} \Pi_{1,1} + \Pi_{2,1},
\qquad
\Pi_{10,10} = \frac{1}{10} \, \Pi_{1,1},
\end{equation}
\begin{equation}
\Pi_{25,24} = \frac{990874363}{356948592}\, B_{1} \Pi_{1,1} + \Pi_{2,1},
\qquad
\Pi_{25,25} = \frac{1}{25}\, \Pi_{1,1},
\end{equation}
where (see~Refs.~\cite{RPert1L, RPert2L, BetaPert1L, BetaPert2L})
\begin{equation}
\Pi_{1,1} = \frac{4}{\beta_{0}},
\qquad
\Pi_{2,1} = \biggl(\frac{4}{\beta_{0}}\biggr)^{\!\!2}
\Biggl\{\frac{365}{24} - 11\zeta(3)
+\nf\!\left[-\frac{11}{12} + \frac{2}{3}\zeta(3)\right]\!\Biggr\},
\end{equation}
\begin{equation}
B_{j} = \frac{\beta_{j}}{\beta_{0}^{j+1}},
\qquad
\beta_{0}=11- \frac{2}{3}\,\nf,
\qquad
\beta_{1} = 102 - \frac{38}{3}\,\nf,
\end{equation}
and~$\zeta(3) \simeq 1.2021$ stands for the Riemann $\zeta$~function.

Additionally, Eq.~(\ref{PIjkOdd}) enables one to obtain, at an arbitrary
loop level, the explicit expression for the
coefficients~$\delta_{j}$~(\ref{DeltaGen}) embodying the contributions of
the~\mbox{$\pi^2$--terms} in the perturbative expression for
the~$R$--ratio of electron--positron annihilation into
hadrons~(\ref{Rappr}), namely~$(\delta_{1}=\delta_{2}=0)$
\begin{align}
\label{DeltaExpl}
\delta_{j} = & -\sum_{k=1}^{K(j)}
\frac{(-2\pi^{2})^{k}}{(2k+1)!}
\sum_{i_{1}=2(k-1)+1}^{j-2}
\underbrace{%
\sum_{i_{2}=2(k-2)+1}^{i_{1}-2}   \ldots
\sum_{i_{n}=2(k-n)+1}^{i_{n-1}-2} \ldots
\sum_{i_{k}=1}^{i_{k-1}-2}%
}_{\mbox{{\small$(k-1)$~sums}}}
(j+i_{1})i_{1} \times
\nonumber \\[1.5mm] & \times
\underbrace{%
(i_{1}+i_{2})i_{2} \times \ldots \times
(i_{n-1}+i_{n})i_{n} \times \ldots \times
(i_{k-1}+i_{k})i_{k}%
}_{\mbox{{\small$(k-1)$~products}}}
\times
\nonumber \\[2mm] & \times
\mathfrak{B}_{j-i_{1}-2}
\underbrace{%
\mathfrak{B}_{i_{1}-i_{2}-2} \ldots
\mathfrak{B}_{i_{n-1}-i_{n}-2} \ldots
\mathfrak{B}_{i_{k-1}-i_{k}-2}%
}_{\mbox{{\small$(k-1)$~terms}}}
d_{i_{k}},
\qquad
d_{j} =\Pi_{j,1},
\qquad
j \ge 3,
\end{align}
where the function~$K(n)$ is defined in Eq.~(\ref{KImDef}) and the
coefficients~$\mathfrak{B}_{n}$ are specified in~Eq.~(\ref{CBdef}), see
also Refs.~\cite{Book, EPJC77, JPG46} and references therein for a
detailed discussion of this issue. It~is straightforward to verify that
Eq.~(\ref{DeltaExpl}) reproduces all the explicit expressions for the
coefficients~$\delta_{j}$ obtained earlier by making use of other methods,
see, in particular, Appendix~C of Ref.~\cite{Book}, Sect.~4 of
Ref.~\cite{EPJC77}, and Sect.~2.3 of Ref.~\cite{JPG46}.

As~mentioned above, the $\pi^{2}$--terms play a valuable role in the
studies of electron--positron annihilation into hadrons and the related
physical observables. Specifically, as~demonstrated in Ref.~\cite{JPG46},
the higher--order \mbox{$\pi^2$--terms}~(\ref{DeltaExpl}) make a
significant effect on the evaluated strong running coupling and the
QCD~scale parameter in the energy range of current BESIII
experiment~\cite{BES}. In~turn, the effect of the higher--order
\mbox{$\pi^2$--terms}~(\ref{DeltaExpl}) on the resulting value of the
strong running coupling in the future ILC~experiment energy
range~\cite{ILC} and the corresponding effect on the \mbox{$R$--ratio}
itself in the future CLIC~experiment energy range~\cite{CLIC} are studied
in Refs.~\cite{JPG46} and~\cite{EPJC77}, respectively.

At the same time, as discussed in Sect.~\ref{Sect:PDRpert}, for a general
choice of the renormalization scale~$\mu^2$ all the higher--order
coefficients~$\Pi_{j,k}$ contribute to the Adler function. Thus, the
obtained results~(\ref{PIj2Rec})--(\ref{PIjkEven}) can also be employed in
the studies of the renormalization scale setting in the theoretical
expressions for the relevant physical observables, such as, e.g., the
$R$--ratio of electron--positron annihilation into hadrons in the energy
range of current BESIII~\cite{BES} and KEDR~\cite{KEDR} experiments.

\section{Conclusions}
\label{Sect:Concl}

The renormalization group relations for the higher--order hadronic vacuum
polarization function perturbative expansion coefficients~$\Pi_{j,k}$ are
studied. The folded recurrent [Eqs.~(\ref{PIj2Rec}),~(\ref{PIjkRec})] and
unfolded explicit [Eqs.~(\ref{PIjkOdd}),~(\ref{PIjkEven})] forms of such
relations are obtained. The explicit expression for the
coefficients~$\delta_{j}$~[Eq.~(\ref{DeltaExpl})] embodying the
contributions of the $\pi^2$--terms in the perturbative expansion of the
$R$--ratio of electron--positron annihilation into hadrons, which play a
substantial role in the studies of the process on~hand, is derived at an
arbitrary loop level. The obtained results can also be employed as an
independent crosscheck of the higher--order perturbative calculations of
the hadronic vacuum polarization function and in the studies of the
renormalization scale setting in the relevant physical observables.

\appendix

\section{RG relations for the coefficients~$\Pi_{j,k}$}
\label{Sect:CoeffsRG}

As discussed in Sect.~\ref{Sect:RR}, at any given order~$j$ the hadronic
vacuum polarization function perturbative expansion
coefficients~$\Pi_{j,k}$ $(j \ge 2; \, k=2,\ldots,j+1)$ entering
Eq.~(\ref{DpertMu}) can be expressed in terms of the
coefficients~$\Pi_{i,1}$~$(i=1,\ldots,j-1)$ by making use of the
RG~equation~(\ref{RGeqnD}). The~corresponding relations for the
coefficients~$\Pi_{j,k}$ at the first eight loop levels are presented in
the following ($\Pi_{j,j+1}=0$ for~$j \ge 1$).

\begin{equation}
\Pi_{j,2} = \frac{1}{2} \sum_{i=1}^{j-1} i\, B_{j-i-1} \Pi_{i,1},
\qquad
\Pi_{j,j} = \frac{1}{j}\,\Pi_{1,1},
\qquad
j \ge 2.
\end{equation}
\begin{equation}
\Pi_{4,3} = \frac{5}{6} B_{1} \Pi_{1,1} + \Pi_{2,1}.
\end{equation}
\begin{equation}
\Pi_{5,3} = \biggl(\frac{1}{2}\, B_{1}^{2} + B_{2} \!\biggr)\Pi_{1,1}
+ \frac{7}{3} B_{1} \Pi_{2,1} + 2 \Pi_{3,1},
\end{equation}
\begin{equation}
\Pi_{5,4} = \frac{13}{12} B_{1} \Pi_{1,1} + \Pi_{2,1}.
\end{equation}
\begin{equation}
\Pi_{6,3} = \frac{7}{6} \biggl( B_{1} B_{2} + B_{3} \!\biggr)\Pi_{1,1}
+ \frac{8}{3} \biggl( \frac{1}{2} B_{1}^2 + B_{2} \!\biggr) \Pi_{2,1}
+ \frac{9}{2} B_{1} \Pi_{3,1} + \frac{10}{3}\Pi_{4,1},
\end{equation}
\begin{equation}
\Pi_{6,4} = \biggl( \frac{35}{24}B_{1}^{2} + \frac{3}{2} B_{2} \!\biggr)\Pi_{1,1}
+ \frac{47}{12} B_{1} \Pi_{2,1} + \frac{5}{2}\Pi_{3,1},
\end{equation}
\begin{equation}
\Pi_{6,5} = \frac{77}{60} B_{1} \Pi_{1,1} + \Pi_{2,1}.
\end{equation}
\begin{align}
\Pi_{7,3} = \, & \frac{4}{3}\biggl( \!B_{1}B_{3} + \frac{1}{2} B_{2}^{2}
+ B_{4} \!\biggr)\Pi_{1,1} + 3 \biggl(\! B_{1} B_{2} + B_{3} \!\biggr)\Pi_{2,1}
+ \nonumber \\[1mm] &
+ 5 \biggl(\frac{1}{2} B_{1}^{2} + B_{2} \!\biggr)\Pi_{3,1}
+ \frac{22}{3} B_{1} \Pi_{4,1} + 5\Pi_{5,1},
\end{align}
\begin{equation}
\Pi_{7,4} = \biggl(\frac{5}{8}B_{1}^{3}+\frac{23}{6}B_{1}B_{2}+2 B_{3}\!\biggr)\Pi_{1,1}
+ \biggl(\frac{59}{12} B_{1}^{2} + 5 B_{2} \biggr)\Pi_{2,1} + \frac{37}{4} B_{1} \Pi_{3,1}
+ 5 \Pi_{4,1},
\end{equation}
\begin{equation}
\Pi_{7,5} = \biggl(\frac{17}{6}B_{1}^{2} + 2 B_{2} \biggr)\Pi_{1,1}
+ \frac{57}{10} B_{1} \Pi_{2,1} + 3\Pi_{3,1},
\end{equation}
\begin{equation}
\Pi_{7,6} = \frac{29}{20} B_{1} \Pi_{1,1} + \Pi_{2,1}.
\end{equation}
\begin{align}
\Pi_{8,3} = \, & \frac{3}{2} \biggl( B_{1} B_{4} + B_{2} B_{3} + B_{5} \biggr)\Pi_{1,1}
+ \frac{10}{3} \biggl(\! B_{1} B_{3} + \frac{1}{2} B_{2}^{2} + B_{4}\!\biggr)\Pi_{2,1}
+ \nonumber \\[1mm] &
+ \frac{11}{2} \biggl( B_{1} B_{2} + B_{3} \biggr)\Pi_{3,1}
+ 8 \biggl( \frac{1}{2} B_{1}^{2} + B_{2} \biggr)\Pi_{4,1}
+ \frac{65}{6} B_{1} \Pi_{5,1} + 7 \Pi_{6,1},
\end{align}
\begin{align}
\Pi_{8,4} = \, & \biggl( \frac{19}{8} B_{1}^{2} B_{2} + \frac{59}{12} B_{1} B_{3} +
\frac{29}{12} B_{2}^{2} + \frac{31}{12} B_{4} \biggr)\Pi_{1,1}
+ \biggl(\! 2 B_{1}^{3} + \frac{73}{6} B_{1} B_{2} + \frac{25}{4} B_{3} \biggr)\Pi_{2,1}
+ \nonumber \\[1mm] &
+ \biggl( \frac{89}{8} B_{1}^{2} + \frac{45}{4} B_{2} \biggr)\Pi_{3,1}
+ \frac{107}{6} B_{1} \Pi_{4,1} + \frac{35}{4}\Pi_{5,1},
\end{align}
\begin{equation}
\Pi_{8,5} =
\biggl(\frac{21}{8} B_{1}^{3} + \frac{33}{4} B_{1} B_{2} + 3 B_{3} \biggr)\Pi_{1,1}
+ \biggl(\frac{139}{12} B_{1}^{2} + 8 B_{2} \biggr)\Pi_{2,1}
+ \frac{319}{20} B_{1} \Pi_{3,1} + 7\Pi_{4,1},
\end{equation}
\begin{equation}
\Pi_{8,6} = \biggl(\frac{413}{90} B_{1}^{2} + \frac{5}{2} B_{2} \!\biggr)\Pi_{1,1}
+ \frac{153}{20} B_{1} \Pi_{2,1} + \frac{7}{2}\Pi_{3,1},
\end{equation}
\begin{equation}
\Pi_{8,7} = \frac{223}{140} B_{1} \Pi_{1,1} + \Pi_{2,1}.
\end{equation}


\begin{thebibliography}{99}


\bibitem{FCC} FCC~Collaboration,
  % A.~Abada~\textit{et al.} [FCC Collaboration],
  % ``FCC Physics Opportunities'',
  Eur.\ Phys.\ J.\ C~\textbf{79}, 474 (2019);
  % report CERN--ACC--2018--0056 (2018);
  % doi:10.1140/epjc/s10052-019-6904-3
  %%CITATION = doi:10.1140/epjc/s10052-019-6904-3;%%
%
  % A.~Abada~\textit{et al.} [FCC Collaboration],
  % ``FCC--ee: The Lepton Collider'',
  Eur.\ Phys.\ J.\ ST~\textbf{228}, 261 (2019);
  % report CERN--ACC--2018--0057 (2018);
  % doi:10.1140/epjst/e2019-900045-4
  %%CITATION = doi:10.1140/epjst/e2019-900045-4;%%
%
  % A.~Abada~\textit{et al.} [FCC Collaboration],
  % ``FCC--hh: The Hadron Collider'',
  % Eur.\ Phys.\ J.\ ST~
  \textbf{228}, 755 (2019);
  % report CERN--ACC--2018--0058 (2018);
  % doi:10.1140/epjst/e2019-900087-0
  %%CITATION = doi:10.1140/epjst/e2019-900087-0;%%
%
  % A.~Abada~\textit{et al.} [FCC Collaboration],
  % ``HE--LHC: The High--Energy Large Hadron Collider'',
  % Eur.\ Phys.\ J.\ ST~
  \textbf{228}, 1109 (2019);
  % doi:10.1140/epjst/e2019-900088-6
  %%CITATION = doi:10.1140/epjst/e2019-900088-6;%%
%
  A.~Blondel~\textit{et al.},
  % ``FCC--ee: Your questions answered'',
  arXiv:1906.02693~[hep-ph].
  %%CITATION = ARXIV:1906.02693;%%


\bibitem{CEPC}
  CEPC Study Group,
  % ``CEPC Conceptual Design Report: Volume 1 -- Accelerator'',
  arXiv:1809.00285~[physics.acc-ph];
  %%CITATION = ARXIV:1809.00285;%%
%
  % CEPC Study Group,
  % ``CEPC Conceptual Design Report: Volume 2 -- Physics and Detector'',
  arXiv:1811.10545~[hep-ex].
  %%CITATION = ARXIV:1811.10545;%%


\bibitem{ILC}
  ILC~Collaboration,
  % T.~Behnke~\textit{et al.} [ILC Collaboration],
  % ``The International Linear Collider Technical Design Report -- Volume~1:
  % Executive Summary'',
  arXiv:1306.6327~[physics.acc-ph];
  %%CITATION = ARXIV:1306.6327;%%
%
  % C.~Adolphsen~\textit{et al.},
  % ``The International Linear Collider Technical Design Report -- Volume~3.II:
  % Accelerator Baseline Design'',
  arXiv:1306.6328~[physics.acc-ph];
  %%CITATION = ARXIV:1306.6328;%%
%
  % T.~Behnke~\textit{et al.},
  % ``The International Linear Collider Technical Design Report - Volume~4:
  % Detectors'',
  arXiv:1306.6329~[physics.ins-det];
  %%CITATION = ARXIV:1306.6329;%%
%
  % H.~Baer \textit{et al.},
  % ``The International Linear Collider Technical Design Report -- Volume~2:
  % Physics'',
  arXiv:1306.6352~[hep-ph];
  %%CITATION = ARXIV:1306.6352;%%
%
  % C.~Adolphsen~\textit{et al.},
  % ``The International Linear Collider Technical Design Report -- Volume~3.I:
  % Accelerator R\&D in the Technical Design Phase'',
  arXiv:1306.6353~[physics.acc-ph].
  %%CITATION = ARXIV:1306.6353;%%


\bibitem{CLIC}
  CLICdp and CLIC Collaborations,
  % P.N.~Burrows~\textit{et al.} [CLICdp and CLIC Collaborations],
  % ``The Compact Linear Collider (CLIC) --- 2018 Summary Report'',
  CERN Yellow Rep.\ Monogr.\ Vol.~2 (2018);
  % doi:10.23731/CYRM-2018-002
  % arXiv:1812.06018~[physics.acc-ph].
  %%CITATION = doi:10.23731/CYRM-2018-002;%%
%
  % J.~de Blas~\textit{et al.},
  % ``The CLIC Potential for New Physics'',
  % CERN Yellow Rep.\ Monogr.\
  Vol.~3 (2018);
  % doi:10.23731/CYRM-2018-003
  % arXiv:1812.02093~[hep-ph].
  %%CITATION = doi:10.23731/CYRM-2018-003;%%
%
  % M.~Aicheler~\textit{et al.} [CLIC accelerator Collaboration],
  % ``The Compact Linear Collider (CLIC) --- Project Implementation Plan'',
  % doi:10.23731/CYRM-2018-004
  % CERN Yellow Rep.\ Monogr.\
  Vol.~4 (2018);
  % arXiv:1903.08655~[physics.acc-ph].
  %%CITATION = doi:10.23731/CYRM-2018-004;%%
%
  % D.~Dannheim~\textit{et al.},
  %``Detector Technologies for CLIC,''
  % CERN Yellow Rep.\ Monogr.\
  Vol.~1 (2019).
  % doi:10.23731/CYRM-2019-001
  % arXiv:1905.02520~[physics.ins-det].
  %%CITATION = doi:10.23731/CYRM-2019-001;%%


\bibitem{E989}
  Muon g--2 Collaboration,
  % J.~Grange \textit{et al.} [Muon g--2 Collaboration],
  % ``Muon (g--2) Technical Design Report'',
  arXiv:1501.06858~[physics.ins-det];
  %%CITATION = ARXIV:1501.06858;%%
%
  % A.~Driutti [Muon g-2 Collaboration],
  % ``Status of the Muon g--2 experiment at Fermilab'',
  SciPost Phys.\ Proc.~\textbf{1}, 033 (2019).
  % doi:10.21468/SciPostPhysProc.1.033
  %%CITATION = doi:10.21468/SciPostPhysProc.1.033;%%


\bibitem{E34}
  E34~Collaboration,
  % M.~Otani [E34~Collaboration],
  % ``Status of the Muon g--2/EDM Experiment at J--PARC (E34)'',
  JPS Conf.\ Proc.~\textbf{8}, 025008 (2015);
  % doi:10.7566/JPSCP.8.025008
  %%CITATION = doi:10.7566/JPSCP.8.025008;%%
%
 Y.~Sato,
  % ``Muon g--2/EDM experiment at J--PARC'',
  PoS~(KMI2017), 006 (2017);
  % doi:10.22323/1.294.0006
  %%CITATION = doi:10.22323/1.294.0006;%%
%
  M.~Abe~\textit{et al.},
  % ``A new approach for measuring the muon anomalous magnetic moment and
  % electric dipole moment'',
  PTEP~\textbf{5}, 053C02 (2019).
  % doi:10.1093/ptep/ptz030
  % arXiv:1901.03047~[physics.ins-det].
  %%CITATION = doi:10.1093/ptep/ptz030;%%


\bibitem{MUonE}
  C.M.~Carloni Calame, M.~Passera, L.~Trentadue, and G.~Venanzoni,
  % ``A~new approach to evaluate the leading hadronic corrections to the muon~$g-2$'',
  Phys.\ Lett.~B~\textbf{746}, 325 (2015);
  % doi:10.1016/j.physletb.2015.05.020
  % arXiv:1504.02228~[hep-ph].
  %%CITATION = doi:10.1016/j.physletb.2015.05.020;%%
%
  G.~Abbiendi~\textit{et al.},
  % ``Measuring the leading hadronic contribution to the muon~$g-2$ via~$\mu e$ scattering'',
  Eur.\ Phys.\ J.~C~\textbf{77}, 139 (2017);
  % doi:10.1140/epjc/s10052-017-4633-z
  % arXiv:1609.08987~[hep-ex].
  %%CITATION = doi:10.1140/epjc/s10052-017-4633-z;%%
%
  P.~Mastrolia, M.~Passera, A.~Primo, and U.~Schubert,
  % ``Master integrals for the NNLO virtual corrections to $\mu e$
  % scattering in QED: the planar graphs'',
  JHEP~\textbf{11}, 198 (2017);
  % doi:10.1007/JHEP11(2017)198
  % arXiv:1709.07435~[hep-ph].
  %%CITATION = doi:10.1007/JHEP11(2017)198;%%
%
  S.~Di Vita, S.~Laporta, P.~Mastrolia, A.~Primo, and U.~Schubert,
  % ``Master integrals for the NNLO virtual corrections to $\mu e$
  % scattering in QED: the non--planar graphs'',
  % JHEP
  \textit{ibid.}~\textbf{09}, 016 (2018);
  % doi:10.1007/JHEP09(2018)016
  % arXiv:1806.08241~[hep-ph].
  %%CITATION = doi:10.1007/JHEP09(2018)016;%%
%
  M.~Fael,
  % ``Hadronic corrections to $\mu$-$e$ scattering at NNLO with space--like data'',
  % JHEP
  \textit{ibid.}~\textbf{02}, 027 (2019);
  % doi:10.1007/JHEP02(2019)027
  % arXiv:1808.08233~[hep-ph].
  %%CITATION = ARXIV:1808.08233;%%
%
  M.~Alacevich~\textit{et~al.},
  % C.M.~Carloni Calame, M.~Chiesa, G.~Montagna, O.~Nicrosini, and F.~Piccinini,
  % ``Muon--electron scattering at NLO'',
  % JHEP
  \textit{ibid.}~\textbf{02}, 155 (2019);
  % doi:10.1007/JHEP02(2019)155
  % arXiv:1811.06743~[hep-ph].
  %%CITATION = doi:10.1007/JHEP02(2019)155;%%
%
  G.~Venanzoni, % [MUonE Collaboration],
  % ``The MUonE experiment: a novel way to measure the leading order
  % hadronic contribution to the muon g-2'',
  PoS~(ICHEP2018), 519 (2019);
  % doi:10.22323/1.340.0519
  % arXiv:1811.11466~[hep-ex].
  %%CITATION = doi:10.22323/1.340.0519;%%
%
  U.~Marconi,
  % ``The MUonE experiment'',
  EPJ Web Conf.~\textbf{212}, 01003 (2019);
  % doi:10.1051/epjconf/201921201003
  %%CITATION = doi:10.1051/epjconf/201921201003;%%
%
  G.~Ballerini \textit{et~al.},
  % G.~Abbiendi, M.~Bonanomi, C.~Brizzolari, U.~Marconi, V.~Mascagna,
  % C.~Matteuzzi, M.~Prest, A.~Principe, M.~Soldani, E.~Vallazza, G.~Venanzoni
  % ``A~feasibility test run for the MUonE project'',
  Nucl.\ Instrum.\ Meth.\ A~\textbf{936}, 636 (2019);
  % doi:10.1016/j.nima.2018.10.148
  %%CITATION = doi:10.1016/j.nima.2018.10.148;%%
%
  M.~Fael and M.~Passera,
  % ``Muon--electron scattering at NNLO: the hadronic corrections'',
  Phys.\ Rev.\ Lett.~\textbf{122}, 192001 (2019);
  % doi:10.1103/PhysRevLett.122.192001
  % arXiv:1901.03106~[hep-ph].
  %%CITATION = doi:10.1103/PhysRevLett.122.192001;%%
%
  P.~Banerjee~\textit{et al.},
  % C.M.~Carloni Calame, M.~Chiesa, S.~Di Vita, T.~Engel, M.~Fael, S.~Laporta,
  % P.~Mastrolia, G.~Montagna, O.~Nicrosini, G.~Ossola, M.~Passera, F.~Piccinini,
  % A.~Primo, J.~Ronca, A.~Signer, W.J.~Torres Bobadilla, L.~Trentadue,
  % Y.~Ulrich and G.~Venanzoni,
  % ``Theory for muon--electron scattering $\@$~10~ppm:
  % A report of the MUonE theory initiative'',
  Eur.\ Phys.\ J.~C~\textbf{80}, 591 (2020);
  % doi:10.1140/epjc/s10052-020-8138-9
  % arXiv:2004.13663~[hep-ph].
%
  A.~Masiero, P.~Paradisi, and M.~Passera,
  % ``New physics at the MUonE experiment at CERN'',
  Phys.\ Rev.\ D~\textbf{102}, 075013 (2020).
  % doi:10.1103/PhysRevD.102.075013
  % arXiv:2002.05418~[hep-ph].


\bibitem{ALEPH}
% \bibitem{ALEPH97}
  R.~Barate~\textit{et al.} [ALEPH~Collaboration],
  % ``Measurement of the spectral functions of
  % vector current hadronic tau decays'',
  Z.\ Phys.\ C~\textbf{76}, 15 (1997);
  % doi:10.1007/s002880050523
%
  % \bibitem{ALEPH98}
  % R.~Barate~\textit{et al.} [ALEPH~Collaboration],
  % ``Measurement of the spectral functions of
  % axial - vector hadronic tau decays and
  % determination of alpha(S)(M**2(tau))'',
  Eur.\ Phys.\ J.\ C~\textbf{4}, 409 (1998);
  % doi:10.1007/s100520050217
%
% \bibitem{ALEPH05}
  S.~Schael~\textit{et al.} [ALEPH Collaboration],
  % ``Branching ratios and spectral functions of tau decays:
  % Final ALEPH measurements and physics implications'',
  Phys.\ Rept.~\textbf{421}, 191 (2005);
  % arXiv:hep-ex/0506072
  % doi:10.1016/j.physrep.2005.06.007
%
  % \bibitem{Davier06}
  M.~Davier, A.~Hocker, and Z.~Zhang,
  % ``The physics of hadronic tau decays'',
  Rev.\ Mod.\ Phys.~\textbf{78}, 1043 (2006);
  % arXiv:hep-ph/0507078.
  %%CITATION = HEP-PH/0507078;%%
%
  % \bibitem{Davier08}
  M.~Davier, S.~Descotes--Genon, A.~Hocker, B.~Malaescu, and Z.~Zhang,
  % ``The determination of alpha(s) from tau decays revisited'',
  Eur.\ Phys.\ J.\ C~\textbf{56}, 305 (2008);
  % arXiv:0803.0979~[hep-ph].
  %%CITATION = ARXIV:0803.0979;%%
%
% \bibitem{Davier14}
  M.~Davier, A.~Hocker, B.~Malaescu, C.~Yuan, and Z.~Zhang,
  % ``Update of the ALEPH non-strange spectral functions from hadronic $\tau$ decays'',
  Eur.\ Phys.\ J.\ C~\textbf{74}, 2803 (2014).
  % arXiv:1312.1501~[hep-ex].
  % doi:10.1140/epjc/s10052-014-2803-9


\bibitem{OPAL}
  % \bibitem{OPAL99}
  K.~Ackerstaff~\textit{et al.}~[OPAL Collaboration],
  % ``Measurement of the strong coupling constant alpha(s)
  % and the vector and axial vector spectral functions in
  % hadronic tau decays'',
  Eur.\ Phys.\ J.\ C~\textbf{7}, 571 (1999);
  % arXiv:hep-ex/9808019.
  % doi:10.1007/s100529901061
%
  % \bibitem{Boito11}
  D.~Boito~\textit{et al.},
  % O.~Cata, M.~Golterman, M.~Jamin, K.~Maltman, J.~Osborne, and S.~Peris,
  % ``A new determination of \alpha_s from hadronic \tau~decays'',
  Phys.\ Rev.\ D~\textbf{84}, 113006 (2011);
  % arXiv:1110.1127~[hep-ph].
  %%CITATION = ARXIV:1110.1127;%%
%
  % \bibitem{Boito12}
  % D.~Boito~\textit{et al.},
  % M.~Golterman, M.~Jamin, A.~Mahdavi, K.~Maltman,
  % J.~Osborne, and S.~Peris,
  % ``An updated determination of $\alpha_s$ from $\tau$ decays'',
  % Phys.\ Rev.\ D~
  \textbf{85}, 093015 (2012);
  % arXiv:1203.3146~[hep-ph].
  %%CITATION = ARXIV:1203.3146;%%
%
% \bibitem{Boito12}
  % D.~Boito~\textit{et al.},
  % M.~Golterman, K.~Maltman, J.~Osborne, and S.~Peris,
  % ``Strong coupling from the revised ALEPH data for hadronic $\tau$~decays'',
  % Phys.\ Rev.\ D~
  \textbf{91}, 034003 (2015).
  % arXiv:1410.3528~[hep-ph].
  % doi:10.1103/PhysRevD.91.034003


\bibitem{HFLAV}
  % \bibitem{HFLAV17}
  Y.~Amhis~\textit{et al.} [HFLAV~Collaboration],
  % ``Averages of $b$-hadron, $c$-hadron, and $\tau$-lepton properties as of summer 2016'',
  Eur.\ Phys.\ J.\ C~\textbf{77}, 895 (2017);
  % doi:10.1140/epjc/s10052-017-5058-4
  % arXiv:1612.07233~[hep-ex].
%
  % \bibitem{HFLAV19}
  % Y.S.~Amhis~\textit{et al.} [HFLAV~Collaboration],
  % ``Averages of $b$-hadron, $c$-hadron, and~$\tau$-lepton properties as of~2018'',
  arXiv:1909.12524~[hep-ex].


\bibitem{BES}
% \bibitem{BES2000}
  J.~Bai~\textit{et al.} [BES~Collaboration],
  % ``Measurement of the total cross-section for hadronic production
  % by e+ e- annihilation at energies between 2.6-GeV - 5-GeV'',
  Phys.\ Rev.\ Lett.~\textbf{84}, 594 (2000);
  % doi:10.1103/PhysRevLett.84.594
  % arXiv:hep-ex/9908046.
%
  % \bibitem{BES2002}
  % J.~Bai~\textit{et al.} [BES~Collaboration],
  % ``Measurements of the cross-section for e+ e ---> hadrons at
  % center-of-mass energies from 2-GeV to 5-GeV'',
  % Phys.\ Rev.\ Lett.~
  \textbf{88}, 101802 (2002);
  % doi:10.1103/PhysRevLett.88.101802
  % arXiv:hep-ex/0102003.
%
  % \bibitem{BES2006}
  M.~Ablikim~\textit{et~al.} [BES~Collaboration],
  % ``Measurements of the cross-sections for e+ e- ---> hadrons at
  % 3.650-GeV, 3.6648-GeV, 3.773-GeV and the branching fraction for
  % psi(3770) ---> non - D anti-D'',
  Phys.\ Lett.~B~\textbf{641}, 145 (2006);
  % doi:10.1016/j.physletb.2006.08.049
  % arXiv:hep-ex/0605105.
%
  % \bibitem{BES2008}
  % M.~Ablikim~\textit{et al.} [BES~Collaboration],
  % ``Determination of the psi(3770), psi(4040), psi(4160)
  % and psi(4415) resonance parameters'',
  % Phys.\ Lett.~B~
  \textbf{660}, 315 (2008);
  % doi:10.1016/j.physletb.2007.11.100
  % arXiv:0705.4500~[hep-ex].
%
  % \bibitem{BES2009}
  % M.~Ablikim~\textit{et al.} [BES~Collaboration],
  % ``R value measurements for e+ e- annihilation at 2.60-GeV, 3.07-GeV and 3.65-GeV'',
  % Phys.\ Lett.~B~
  \textbf{677}, 239 (2009);
  % doi:10.1016/j.physletb.2009.05.055
  % arXiv:0903.0900~[hep-ex].
%
  % \bibitem{BES2020}
  % M.~Ablikim~\textit{et al.} [BES~Collaboration],
  % ``Future physics programme of~BESIII'',
  Chin.\ Phys.\ C~\textbf{44}, 040001 (2020).
  % doi:10.1088/1674-1137/44/4/040001
  % arXiv:1912.05983~[hep-ex].


\bibitem{KEDR}
  % \bibitem{KEDR2016}
  V.~Anashin~\textit{et al.} [KEDR~Collaboration],
  % ``Measurement of~$R_{\text{uds}}$ and~$R$ between~3.12
  % and 3.72~GeV at the KEDR detector'',
  Phys.\ Lett.~B~\textbf{753}, 533 (2016);
  % doi:10.1016/j.physletb.2015.12.059
  % arXiv:1510.02667~[hep-ex].
%
  % \bibitem{KEDR2017}
  % V.~Anashin~\textit{et al.} [KEDR~Collaboration],
  % ``Measurement of~$R$ between~1.84 and 3.05~GeV at the KEDR detector'',
  % Phys.\ Lett.~B~
  \textbf{770}, 174 (2017);
  % doi:10.1016/j.physletb.2017.04.073
  % arXiv:1610.02827~[hep-ex].
%
  % \bibitem{KEDR2019}
  % V.~Anashin~\textit{et al.} [KEDR~Collaboration],
  % ``Precise measurement of~$R_{\text{uds}}$ and~$R$ between~1.84
  % and 3.72~GeV at the KEDR detector'',
  % Phys.\ Lett.~B~
  \textbf{788}, 42 (2019).
  % doi:10.1016/j.physletb.2018.11.012
  % arXiv:1805.06235~[hep-ex].


\bibitem{FCCtheory19}
  A.~Blondel~\textit{et al.},
  % ``Theory for the FCC-ee'',
  CERN Yellow Reports: Monographs, CERN--2020--003 (CERN, Geneva, 2020);
  % doi:10.23731/CYRM-2020-003
  arXiv:1905.05078~[hep-ph].


\bibitem{MuonWP20}
  T.~Aoyama~\textit{et al.},
  % ``The anomalous magnetic moment of the muon in the Standard Model'',
  Phys.\ Rept.~\textbf{887}, 1 (2020).
  % doi:10.1016/j.physrep.2020.07.006
  % arXiv:2006.04822~[hep-ph].


\bibitem{Adler}
  S.L.~Adler,
  % ``Some simple vacuum--polarization phenomenology: $\epem\!\!\to\:$hadrons;
  % the muonic--atom $X$--ray discrepancy and ($g_{\mu}-2$)'',
  Phys.\ Rev.\ D~\textbf{10}, 3714 (1974).
  % doi:10.1103/PhysRevD.10.3714
  %%CITATION = doi:10.1103/PhysRevD.10.3714;%%


\bibitem{Rad82}
  A.V.~Radyushkin,
  % ``Optimized $\Lambda$--parametrization for the QCD running
  % coupling constant in spacelike and timelike regions'',
  report JINR E2--82--159 (1982); JINR Rapid Commun.~\textbf{78}, 96 (1996);
  arXiv:hep-ph/9907228.
  %%CITATION = HEP-PH/9907228;%%


\bibitem{KP82}
  N.V.~Krasnikov and A.A.~Pivovarov,
  % ``The influence of the analytic continuation effects on the value of the QCD scale
  % parameter $\Lambda$ extracted from charmonium and upsilonium hadron decays'',
  Phys.\ Lett.\ B~\textbf{116}, 168 (1982).
  % doi:10.1016/0370-2693(82)91001-2
  %%CITATION = doi:10.1016/0370-2693(82)91001-2;%%


\bibitem{Book}
  A.V.~Nesterenko, \textit{Strong interactions in spacelike and timelike
  domains: Dispersive approach}, Elsevier, Amsterdam, 222~p.~(2017).
  %%CITATION = INSPIRE-1514185;%%


\bibitem{RPert1L}
  T.~Appelquist and H.~Georgi,
  % ``$\epem$~annihilation in gauge theories of strong interactions'',
  Phys.\ Rev.\ D~\textbf{8}, 4000 (1973);
  % doi:10.1103/PhysRevD.8.4000
  %%CITATION = doi:10.1103/PhysRevD.8.4000;%%
%
  A.~Zee,
  % ``Electron--positron annihilation in stagnant field theories'',
  % Phys.\ Rev.\ D
  \textit{ibid.}~\textbf{8}, 4038 (1973).
  % doi:10.1103/PhysRevD.8.4038
  %%CITATION = doi:10.1103/PhysRevD.8.4038;%%


\bibitem{Penn}
  R.G.~Moorhouse, M.R.~Pennington, and G.G.~Ross,
  % ``What can asymptotic freedom say about $\epem\to$~hadrons?'',
  Nucl.\ Phys.\ B~\textbf{124}, 285 (1977);
  % doi:10.1016/0550-3213(77)90316-9
  %%CITATION = doi:10.1016/0550-3213(77)90316-9;%%
%
  M.R.~Pennington and G.G.~Ross,
  % ``Perturbative QCD for timelike processes: what is the best expansion parameter?'',
  Phys.\ Lett.\ B~\textbf{102}, 167 (1981);
  % doi:10.1016/0370-2693(81)91055-8
  %%CITATION = doi:10.1016/0370-2693(81)91055-8;%%
%
  M.R.~Pennington, R.G.~Roberts, and G.G.~Ross,
  % ``How to continue the predictions of perturbative QCD from the spacelike region
  % where they are derived to the timelike regime where experiments are performed'',
  Nucl.\ Phys.\ B~\textbf{242}, 69 (1984).
  % doi:10.1016/0550-3213(84)90134-2
  %%CITATION = doi:10.1016/0550-3213(84)90134-2;%%


\bibitem{Pivovarov91}
  A.A.~Pivovarov,
  % ``Renormalization group summation of perturbative series in timelike momentum region'',
  Nuovo Cim.~A~\textbf{105}, 813 (1992).
  % doi:10.1007/BF02799096
  %%CITATION = doi:10.1007/BF02799096;%%


\bibitem{JPG42}
  A.V.~Nesterenko,
  % ``Hadronic vacuum polarization function within dispersive approach to QCD'',
  J.~Phys.\ G~\textbf{42}, 085004 (2015).
  % doi:10.1088/0954-3899/42/8/085004
  % arXiv:1411.2554~[hep-ph].
  %%CITATION = doi:10.1088/0954-3899/42/8/085004;%%


\bibitem{JPG46}
  A.V.~Nesterenko,
  % ``Explicit form of the R--ratio of electron--positron annihilation into hadrons'',
  J.~Phys.\ G~\textbf{46}, 115006 (2019).
  % doi:10.1088/1361-6471/ab433e
  % arXiv:1902.06504~[hep-ph].
  %%CITATION = doi:10.1088/1361-6471/ab433e;%%


\bibitem{RPert4L}
  P.A.~Baikov, K.G.~Chetyrkin, and J.H.~Kuhn,
  % ``Order $\alpha_{s}^{4}$ QCD corrections to~$Z$ and~$\tau$ decays'',
  Phys.\ Rev.\ Lett.~\textbf{101}, 012002 (2008);
  % doi:10.1103/PhysRevLett.101.012002
  % arXiv:0801.1821~[hep-ph].
  %%CITATION = doi:10.1103/PhysRevLett.101.012002;%%
%
  % P.A.~Baikov, K.G.~Chetyrkin, and J.H.~Kuhn,
  % ``Adler function, Bjorken sum rule, and the Crewther relation
  % to order~$\alpha_{s}^{4}$ in a general gauge theory'',
  % Phys.\ Rev.\ Lett.~
  \textbf{104}, 132004 (2010);
  % doi:10.1103/PhysRevLett.104.132004
  % arXiv:1001.3606~[hep-ph].
  %%CITATION = doi:10.1103/PhysRevLett.104.132004;%%
%
  P.A.~Baikov, K.G.~Chetyrkin, J.H.~Kuhn, and J.~Rittinger,
  % ``Adler function, sum rules and Crewther relation of order
  % $\co(\alpha_{s}^{4})$: the singlet case'',
  Phys.\ Lett.\ B~\textbf{714}, 62 (2012).
  % doi:10.1016/j.physletb.2012.06.052
  % arXiv:1206.1288~[hep-ph].
  %%CITATION = doi:10.1016/j.physletb.2012.06.052;%%


\bibitem{BCK0912}
  P.A.~Baikov, K.G.~Chetyrkin, and J.H.~Kuhn,
  % ``$R(s)$ and hadronic $\tau$--decays in order $\alpha^4(s)$: technical aspects'',
  Nucl.\ Phys.\ B (Proc.\ Suppl.) \textbf{189}, 49 (2009);
  % doi:10.1016/j.nuclphysbps.2009.03.010
  % arXiv:0906.2987~[hep-ph].
  %%CITATION = doi:10.1016/j.nuclphysbps.2009.03.010;%%
%
  % \bibitem{BCKR2012}
  P.A.~Baikov, K.G.~Chetyrkin, J.H.~Kuhn, and J.~Rittinger,
  % ``Vector correlator in massless QCD at order $O(alpha_{s}^{4})$ and
  % the QED beta-function at five loop'',
  JHEP \textbf{07}, 017 (2012).
  % doi:10.1007/JHEP07(2012)017
  % arXiv:1206.1284~[hep-ph].
  %%CITATION = doi:10.1007/JHEP07(2012)017;%%


\bibitem{Beta5L}
  P.A.~Baikov, K.G.~Chetyrkin, and J.H.~Kuhn,
  % ``Five--loop running of the QCD coupling constant'',
  Phys.\ Rev.\ Lett.~\textbf{118}, 082002 (2017);
  % doi:10.1103/PhysRevLett.118.082002
  % arXiv:1606.08659~[hep-ph].
  %%CITATION = doi:10.1103/PhysRevLett.118.082002;%%
%
  F.~Herzog, B.~Ruijl, T.~Ueda, J.A.M.~Vermaseren, and A.~Vogt,
  % ``The five--loop beta function of Yang--Mills theory with fermions'',
  JHEP~\textbf{02}, 090 (2017);
  % doi:10.1007/JHEP02(2017)090
  % arXiv:1701.01404~[hep-ph].
  %%CITATION = doi:10.1007/JHEP02(2017)090;%%
%
  T.~Luthe, A.~Maier, P.~Marquard, and~Y.~Schroder,
  % ``The five--loop beta function for a general gauge group and
  % anomalous dimensions beyond Feynman gauge'',
  % JHEP
  \textit{ibid.}~\textbf{10}, 166 (2017).
  % doi:10.1007/JHEP10(2017)166
  % arXiv:1709.07718~[hep-ph].
  %%CITATION = doi:10.1007/JHEP10(2017)166;%%


\bibitem{EPJC77}
  A.V.~Nesterenko,
  % ``Electron--positron annihilation into hadrons at the higher--loop levels'',
  Eur.\ Phys.\ J.\ C~\textbf{77}, 844 (2017).
  % doi:10.1140/epjc/s10052-017-5405-5
  % arXiv:1707.00668~[hep-ph].
  %%CITATION = doi:10.1140/epjc/s10052-017-5405-5;%%


\bibitem{Bj89}
  J.D.~Bjorken,
  % ``Two~topics in Quantum Chromodynamics'',
  report SLAC--PUB--5103 (1989).
  %%CITATION = SLAC-PUB-5103;%%


\bibitem{ProsperiAlpha}
  G.M.~Prosperi, M.~Raciti, and C.~Simolo,
  % ``On~the running coupling constant in~QCD'',
  Prog.\ Part.\ Nucl.\ Phys.~\textbf{58}, 387 (2007).
  % doi:10.1016/j.ppnp.2006.09.001
  % arXiv:hep-ph/0607209.
  %%CITATION = doi:10.1016/j.ppnp.2006.09.001;%%


\bibitem{RPert2L}
  K.G.~Chetyrkin, A.L.~Kataev, and F.V.~Tkachov,
  % ``Higher--order corrections to $\sigma_{{\rm tot}}$($\epem\!\!\to\:$hadrons)
  % in Quantum Chromodynamics'',
  Phys.\ Lett.\ B~\textbf{85}, 277 (1979);
  % doi:10.1016/0370-2693(79)90596-3
  %%CITATION = doi:10.1016/0370-2693(79)90596-3;%%
%
  M.~Dine and J.R.~Sapirstein,
  % ``Higher--order Quantum Chromodynamic corrections in $\epem$~annihilation'',
  Phys.\ Rev.\ Lett.~\textbf{43}, 668 (1979);
  % doi:10.1103/PhysRevLett.43.668
  %%CITATION = doi:10.1103/PhysRevLett.43.668;%%
%
  W.~Celmaster and R.J.~Gonsalves,
  % ``Analytic calculation of higher--order Quantum Chromodynamic
  % corrections in $\epem$~annihilation'',
  % Phys.\ Rev.\ Lett.
  \textit{ibid.}~\textbf{44}, 560 (1980).
  % doi:10.1103/PhysRevLett.44.560
  %%CITATION = doi:10.1103/PhysRevLett.44.560;%%


\bibitem{BetaPert1L}
  D.J.~Gross and F.~Wilczek,
  % ``Ultraviolet behavior of nonabelian gauge theories'',
  Phys.\ Rev.\ Lett.~\textbf{30}, 1343 (1973);
  % doi:10.1103/PhysRevLett.30.1343
  %%CITATION = doi:10.1103/PhysRevLett.30.1343;%%
%
  H.D.~Politzer,
  % ``Reliable perturbative results for strong interactions?'',
  % Phys.\ Rev.\ Lett.
  \textit{ibid.}~\textbf{30}, 1346 (1973).
  % doi:10.1103/PhysRevLett.30.1346
  %%CITATION = doi:10.1103/PhysRevLett.30.1346;%%


\bibitem{BetaPert2L}
  W.E.~Caswell,
  % ``Asymptotic behavior of nonabelian gauge theories to two loop order'',
  Phys.\ Rev.\ Lett.~\textbf{33}, 244 (1974);
  % doi:10.1103/PhysRevLett.33.244
  %%CITATION = doi:10.1103/PhysRevLett.33.244;%%
%
  D.R.T.~Jones,
  % ``Two loop diagrams in Yang--Mills theory'',
  Nucl.\ Phys.\ B~\textbf{75}, 531 (1974);
  % doi:10.1016/0550-3213(74)90093-5
  %%CITATION = doi:10.1016/0550-3213(74)90093-5;%%
%
  E.~Egorian and O.V.~Tarasov,
  % ``Two Loop renormalization of the QCD in an arbitrary gauge'',
  % Teor.\ Mat.\ Fiz.~\textbf{41}, 26 (1979)
  Theor.\ Math.\ Phys.~\textbf{41}, 863 (1979).
  %%CITATION = TMFZA,41,26;%%


\end{thebibliography}
\end{document}